\documentclass[rsi,twocolumn,preprintnumbers,amsmath,amssymb,superscriptaddress,floatfix]{revtex4}

\usepackage{graphicx}% Include figure files
\usepackage{dcolumn}% Align table columns on decimal point
\usepackage{bm}% bold math
\usepackage{here} 
\usepackage{epsf}
\usepackage{epstopdf}
\usepackage{color}

\begin{document}

%\preprint{preprint}
%\preprint{manuscript}

\title{Photoemission System with Polarized Hard X-rays\\for Probing Ground State Symmetry of Strongly Correlated Materials}

%----------------------------------------------- Author ---------------------------------------------------------

\author{H. Fujiwara}
\email{fujiwara@mp.es.osaka-u.ac.jp}
\affiliation{Graduate School of Engineering Science, Osaka University, Toyonaka, Osaka 560-8531, Japan}
\affiliation{SPring-8/RIKEN, Sayo, Hyogo 679-5148, Japan}
\author{S. Naimen} 
\affiliation{Graduate School of Engineering Science, Osaka University, Toyonaka, Osaka 560-8531, Japan}
\author{A. Higashiya}
\affiliation{SPring-8/RIKEN, Sayo, Hyogo 679-5148, Japan}
\affiliation{Faculty of Science and Engineering, Setsunan University, Neyagawa, Osaka 572-8508, Japan}
\author{Y. Kanai}
\author{H. Yomosa}
\author{K. Yamagami}
\author{T. Kiss}
\affiliation{Graduate School of Engineering Science, Osaka University, Toyonaka, Osaka 560-8531, Japan}
\affiliation{SPring-8/RIKEN, Sayo, Hyogo 679-5148, Japan}
\author{T. Kadono}
\author{S. Imada}
\affiliation{SPring-8/RIKEN, Sayo, Hyogo 679-5148, Japan}
\affiliation{Department of Physical Science, Ritsumeikan University, Kusatsu, Shiga 525-8577, Japan}
\author{A. Yamasaki}
\affiliation{SPring-8/RIKEN, Sayo, Hyogo 679-5148, Japan}
\affiliation{Faculty of Science and Engineering, Konan University, Kobe 658-8501, Japan}
\author{K. Takase}
\affiliation{Department of Physics, College of Science and Technology, Nihon University, Chiyoda, Tokyo 101-0062, Japan}
\author{S. Otsuka}
\author{T. Shimizu}
\author{S. Shingubara}
\affiliation{Graduate School of Science and Technology, Kansai University, Suita, Osaka 564-8680, Japan}
\author{S. Suga} 
\affiliation{SPring-8/RIKEN, Sayo, Hyogo 679-5148, Japan}
\affiliation{Institute of Scientific and Industrial Research, Osaka University, Ibaraki, Osaka 567-0047, Japan}
\author{M. Yabashi}
\affiliation{SPring-8/RIKEN, Sayo, Hyogo 679-5148, Japan}
\author{K. Tamasaku}
\affiliation{SPring-8/RIKEN, Sayo, Hyogo 679-5148, Japan}
\author{T. Ishikawa}
\affiliation{SPring-8/RIKEN, Sayo, Hyogo 679-5148, Japan}
\author{A. Sekiyama}
\affiliation{Graduate School of Engineering Science, Osaka University, Toyonaka, Osaka 560-8531, Japan}
\affiliation{SPring-8/RIKEN, Sayo, Hyogo 679-5148, Japan}
\date{\today}

%-----------------abstract -------------------
\begin{abstract}
We have developed a polarized hard X-ray photoemission (HAXPES) system to study the ground-state symmetry of strongly correlated materials. The linear polarization of the incoming X-ray beam is switched by the transmission-type phase retarder composed of two diamond (100) crystals. The best degree of the linear polarization $P_L$ is $-0.96$, containing the vertical polarization component of 98\%. A newly developed low temperature two-axis manipulator enables easy polar and azimuthal rotations to select the detection direction of photoelectrons. The lowest temperature achieved is 9 K, offering us a chance to access the ground state even for the strongly correlated electron systems in cubic symmetry. The co-axial sample monitoring system with the long-working-distance microscope enables us to keep measuring the same region on the sample surface before and after rotation procedures. Combining this sample monitoring system with a micro-focused X-ray beam by means of an ellipsoidal Kirkpatrick-Baez mirror (25 $\mu$m $\times$ 25 $\mu$m (FWHM)), we have demonstrated the polarized valence-band HAXPES on NiO for voltage application as resistive random access memories to reveal the origin of the metallic spectral weight near the Fermi level.
\end{abstract}
%------------------------------------------------

%%%%%%%%

%------------PACS and KEY WORD-------
%\pacs{}% PACS, the Physics and Astronomy     PACS
                             % Classification Scheme.
%\keywords{Suggested keywords}%Use showkeys  class option if keyword
                              %display desired
%------------------------------
\maketitle
%
%------------StartBody------

\section{INTRODUCTION}

Strongly correlated electron systems show variety of anomalous phenomena such as the high T$_c$ superconductivity and the metal-insulator transitions, attracting much attentions in the solid state physics. In the case of the transition-metal $3d$-electron systems, the ground state symmetry can easily be determined due to the crystal-field splitting with the eV-order. However, the situation is different in the case of rare-earth $4f$-electron systems in which the crystal field splitting is usually of the meV-order. Therefore, the determination of the ground state symmetry for the $4f$-electron systems is still a non-trivial issue although it is crucial to discuss the origin of the intriguing phenomena such as quantum criticality, heavy-Fermion superconductivity, and non-Fermi liquid behavior. For this purpose, the linear polarization dependence of the soft-X-ray absorption (XAS) spectra has been well established as a local probe to detect the symmetry of the electronic structure due to the excitonic excitation forming a core hole in the final states with the dipole selection rules. This technique is powerful for the Ce compounds with the tetragonal or orthorhombic crystal structures~\cite{Hansmann2008,Willers2009,Willers2010,Willers2012,Strigari2012,Strigari2013,Willers2014}, while it does not work for a high symmetric cubic system because the selection rule is only sensitive to the case that the electric field vector of the excitation photon is parallel or perpendicular to the anisotropic crystal axis. This fundamental limitation is, however, overcome by a recent development of \textit{`polarized'} hard X-ray photoemission spectroscopy (HAXPES)~\cite{Mori2014}, in which the dipole selection rules equally work as well as the polarized XAS. Moreover we have another controllable probing parameter as a photoelectron-detection direction, enabling to probe the ground state symmetry even for the cubic system~\cite{Kanai2015}.

In this paper, we report on the performance of a HAXPES system with a potential of polarization switching in SPring-8 BL19LXU. The double phase retarders can switch the polarization of the incoming photon from the horizontal to vertical polarization with the degree of the linear polarization $P_L$ of $-0.96$. A two-axis manipulator with the lowest available temperature of 9 K is newly developed to easily select the photoelectron-detection direction in the ground state of strongly correlated electron systems. Moreover, the sample monitoring system with a co-axial long-working-distance microscope enables one to keep measuring the same sample region after rotation procedures. Combined with the microfocus X-ray beam with a size of 25 $\mu$m $\times$ 25 $\mu$m (FWHM), we further demonstrate the linear polarization dependence of the valence-band HAXPES spectra for the voltage-applied NiO to reveal the origin of the resistive switching phenomena.

\section{HAXPES System with Linear Polarization Switching}
\subsection{Beamline Setup and Double-Crystal Phase Retarder}	

\begin{figure}
\begin{center}
\includegraphics[width=8cm,clip]{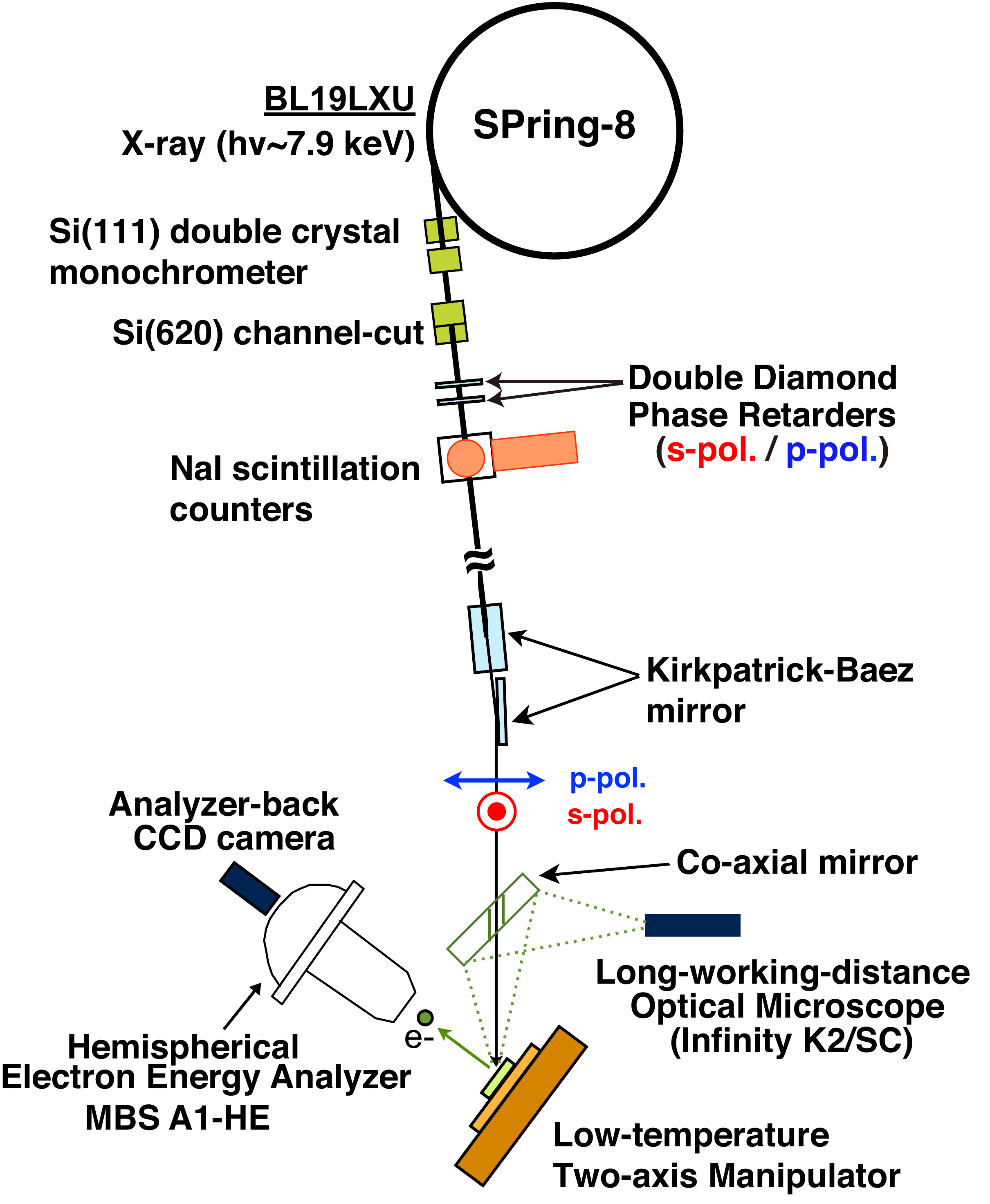}
\caption 
{(color online) Overview of the experimental geometry (top view) for the polarized HAXPES at BL19LXU in SPring-8.}
\label{fig1}
\end{center}
\end{figure}

\begin{figure}
\begin{center}
\includegraphics[width=8cm,clip]{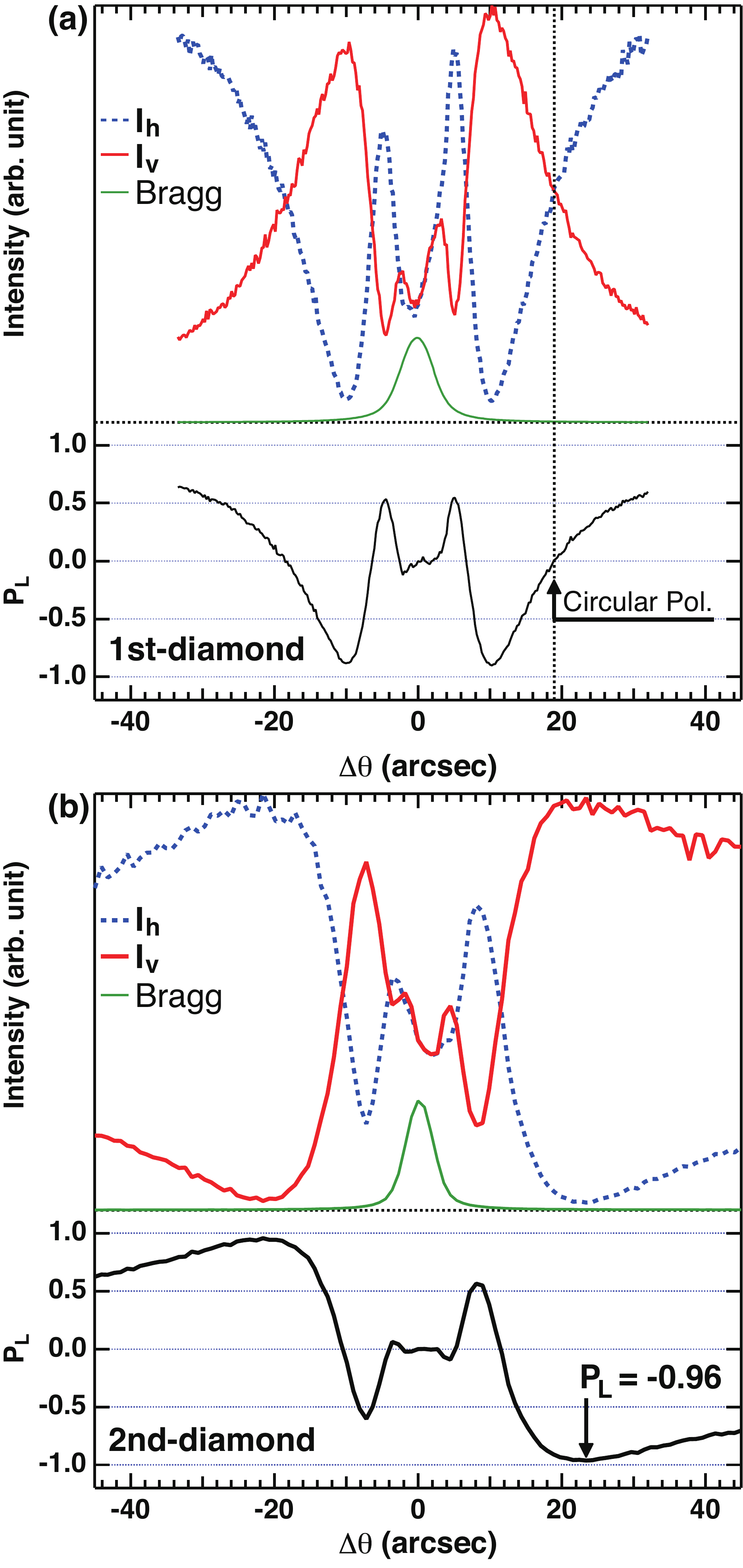}
\caption 
{(color online) Offset angle dependence of the photon intensities relative to the diamond (220) Bragg reflections for the first and second diamond in (a) and (b), respectively. The evaluated $P_L$ (degree of linear polarization) is also plotted in the bottom of (a) and (b).}
\label{fig2}
\end{center}
\end{figure}

The polarization dependent HAXPES was performed at BL19LXU in SPring-8 by using an MBS A1-HE hemispherical photoelectron spectrometer. Figure~\ref{fig1} shows the schematic picture of the typical setting of the  beam-line optics for the polarized HAXPES. The linearly polarized light is delivered from an in-vacuum 27-m long 780 periods undulator~\cite{Kitamura2001,Yabashi2001a,Yabashi2001b}. The photon energy is set to $\sim$7.9 keV by the Si(111) double-crystal monochrometer and further monochromated by the Si(620) channel-cut crystal. The transmission-type phase retarder made of two single-crystalline (100) diamonds are placed at the downstream of the channel-cut crystal to switch the polarization of the X-ray beam~\cite{Hirano1991,Hirano1993,Giles1994,Hirano1997}. To compensate the phase-shift inhomogeneity due to the angular divergence of the incoming X-ray beam, we use the double-phase-plate configuration~\cite{Okitsu2001}, giving the (220) Bragg reflection in opposite directions with Laue geometry, whose scattering plane inclined to 45 degree to the electric field vector of the incoming X-ray beam. The original horizontally polarized photons are transformed into circular polarization by the first diamond plate [Fig.~\ref{fig2}(a)], and then the offset angle of the second plate is scanned as shown in Fig.~\ref{fig2}(b) to give the further phase shift for converting to the vertical polarization~\cite{Scagnoli2009}. The degree of the linear polarization is defined as, 
\begin{equation}
\label{PL}
P_L  = \frac{I_{h}-I_{v}}{I_{h}+I_{v}},
\end{equation}
where $I_{h}$ and $I_{v}$ are intensities for the horizontal and vertical polarized photons detected by NaI scintillation counters located at the downstream of the phase retarder. In our set up, the best $P_L$ value is evaluated as $-0.96$ as shown in Fig.~\ref{fig2}(b). This gives the fraction of the vertical components of 98\%, having enough quality to discuss the linear polarization dependence of the photoemission spectra. Note that the electric field vector of horizontal (vertical) polarized photons is parallel (perpendicular) to the photoelectron scattering plane, and thus defined as $p$($s$)-polarization, respectively. The thickness of each diamond plate is 0.25 mm, and the transmittance of the X-ray beam after the double-crystal phase retarder is $\sim$50\%, which gives much better throughput for the vertically polarized photons than that for a 0.7mm-thick single phase-retarder setup with transmittance of $\sim$35\%~\cite{Sekiyama2010,Nakatsu2011,Sekiyama2013}. Afterwards the X-ray beam is focused onto the sample within 25 ${\mu}$m ${\times}$ 25 ${\mu}$m (FWHM) by the ellipsoidal Kirkpatrick-Baez mirror as shown in Fig.~\ref{fig1}. 

%------------\section{Fig1_SchematicPicture}----------------

%------------\section{Fig2_Diamond_Profile}----------------

\subsection{Low-Temperature Two-axis Manipulator}

%------------\section{Fig3_PHOTO}----------------
\begin{figure*}
\begin{center}
\includegraphics[width=17cm,clip]{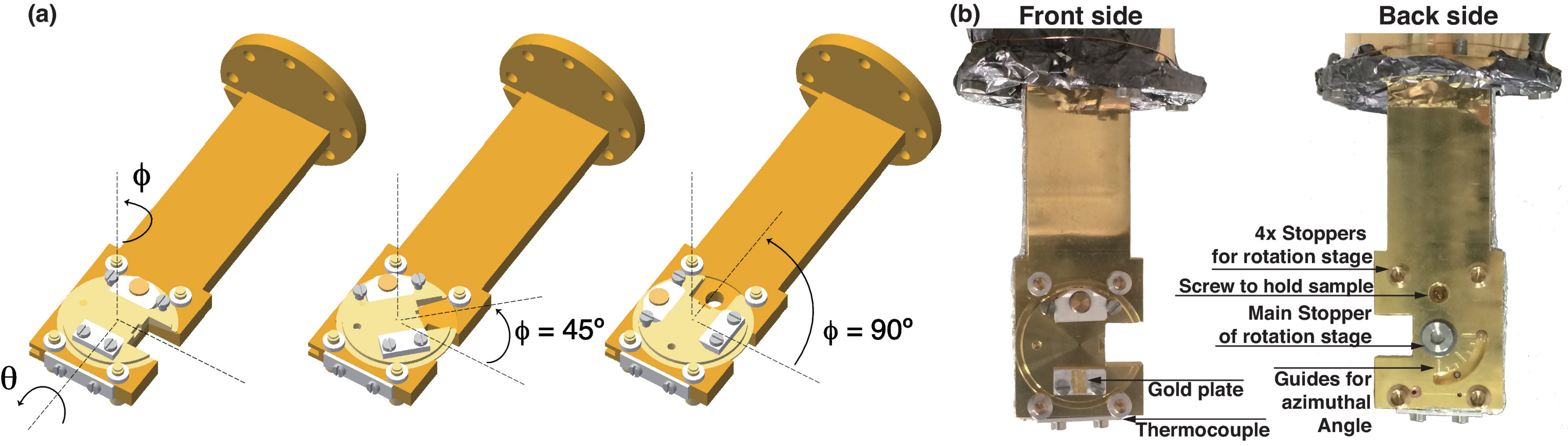}
\caption 
{(color online)(a) Schematic view of our developed low-temperature two-axis manipulator with the definition of azimuthal angle $\phi$ and polar angle $\theta$, and simulations for the azimuthal rotation with rotation angle of 45$^{\circ}$ (middle) and 90$^{\circ}$ (right).(b) Photographs of the manipulator taken from front (left) and back (right) sides.} 
\label{fig3}
\end{center}
\end{figure*}

%------------\section{Fig4_analyzer back CCD and Spectra}----------------
\begin{figure*}
\begin{center}
\includegraphics[width=17cm,clip]{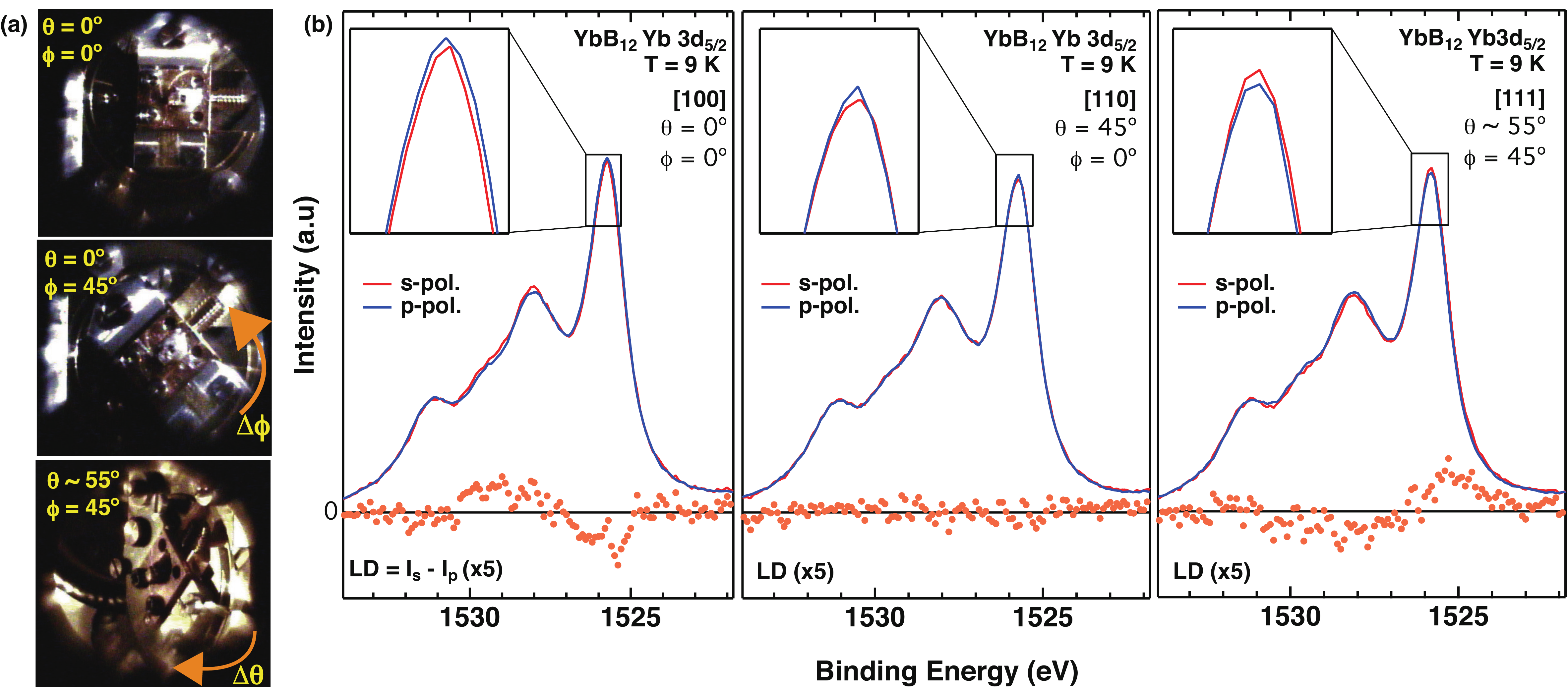}
\caption 
{(color online) (a) CCD camera image of the manipulator (top) with demonstration of the 45$^{\circ}$ rotation for the azimuthal axis (middle) and the further polar rotation of $\sim$55$^{\circ}$(bottom). (b) Linear polarization dependence of the Yb$^{3+}$ $3d_{5/2}$ core level spectra for cubic YbB$_{12}$ by selecting the photoelectron-detection direction of [100] (left), [110] (middle), and [111] (right). Insets are the close-up of 1525.5 eV peak in the Yb$^{3+}$ $3d_{5/2}$ spectra. So-called Shirly-type background is subtracted as discussed in the literature~\cite{Kanai2015}.} 
\label{fig4}
\end{center}
\end{figure*}

We have developed a two-axis manipulator for the polarized HAXPES to easily optimize the detection direction of photoelectrons. Figure~\ref{fig3}(a) shows the schematic drawing of the manipulator. The body of the manipulator is made of the oxygen-free copper with the gold plating. The rotation feedthrough provides the polar rotation ($\theta$), and the rotation stage made of Be-Cu on top of the manipulator gives the azimuthal rotation ($\phi$) in the range of $90^{\circ}$. The bottom of the rotation stage is bowl-shaped to gain the thermal contact, and tightly fixed by the non-magnetic screws from the back side. Furthermore four aluminum stoppers supports the rotation stage from the front side as shown in the photograph in Fig.~\ref{fig3}(b). By minimizing volume of the manipulator, we have achieved the lowest temperature of 9 K by a closed-cycle He refrigerator, enabling to cool down to 9 K in 1.5 hours from 300 K. The sample is mounted on the rotation stage by clamping in-between two aluminum fingers with a screw to hold the sample by the hexagonal wrench from the backside, and the same wrench can be used to manipulate the rotation stage. To guide the rotation angle, there are 7 indications with every 15 degree.

The rotation angle is fixed more precisely by monitoring the sample with a charge-coupled device (CCD) camera, which is mounted on the backside of the hemispherical analyzer. The camera image is captured and shown on the computer screen by using the program with the function to calculate the rotation angle. Figure~\ref{fig4}(a) shows the rotation procedures of the two-axis manipulator recorded by this CCD camera with magnification of $\sim$3 achieved by a varifocal lens (Tamron : 13VM20100AS).  In the middle of Fig.~\ref{fig4}(a), one can clearly find the azimuthal rotation of 45$^{\circ}$, and the further polar rotation with $\sim$55$^{\circ}$ (bottom). 

By selecting the photoelectron-detection direction with the two-axis manipulator, we have studied the polarized HAXPES for a cubic strongly correlated rare-earth compound YbB$_{12}$~\cite{Kanai2015}. Figure~\ref{fig4}(b) shows the Yb$^{3+}$ $3d_{5/2}$ core-level spectra along the [100], [110], and [111] directions. Note that all three geometries have been set by the polar $\theta$ and azimuthal $\phi$ rotations without any sample transfer procedure. In the [100] emission geometry of ($\theta$, $\phi$) = ($0^{\circ}$,$0^{\circ}$), the Yb$^{3+}$ $3d$ core level spectra shows the polarization dependence especially at the peak at 1525.5 eV. As shown in the close-up around the peak-top (inset), the $p$-polarized spectral weight is slightly stronger than the $s$-polarized spectral weight. This is highlighted by the linear dichroism (LD) in the core-level spectra (bottom) defined as $I_{s} - I_{p}$, where $I_s$ ($I_p$) denotes the intensity of $s$-($p$-)polarized spectra. In the [100] emission geometry, the LD spectrum changes the sign near 1527 eV. This difference is noticeably suppressed when one sets to the [110] emission geometry by rotating the polar angle $\theta$ of 45$^{\circ}$, and again shows up with the [111] emission geometry after the rotation procedures for both $\theta$ and $\phi$ axis of ($\theta$, $\phi$) = ($\sim$55$^{\circ}$, $45^{\circ}$), setting in Fig.~\ref{fig4}(a). Most interesting is that the sign of LD for the [111] emission geometry is reversed to that for the [100] emission, revealing the ground state symmetry of the Yb$^{3+}$ $4f$ states due to the Coulomb and exchange interactions between the Yb $3d$ core and $4f$ holes~\cite{Mori2014, Kanai2015}. Thus we stress that the selection of the photoelectron-detection direction is crucial to probe the ground state symmetry for various strongly correlated electron systems.  

\subsection{Co-axial monitoring system with long working-distance microscope}

%------------\section{Fig5_K2_PHOTO}----------------
\begin{figure}
\begin{center}
\includegraphics[width=8cm,clip]{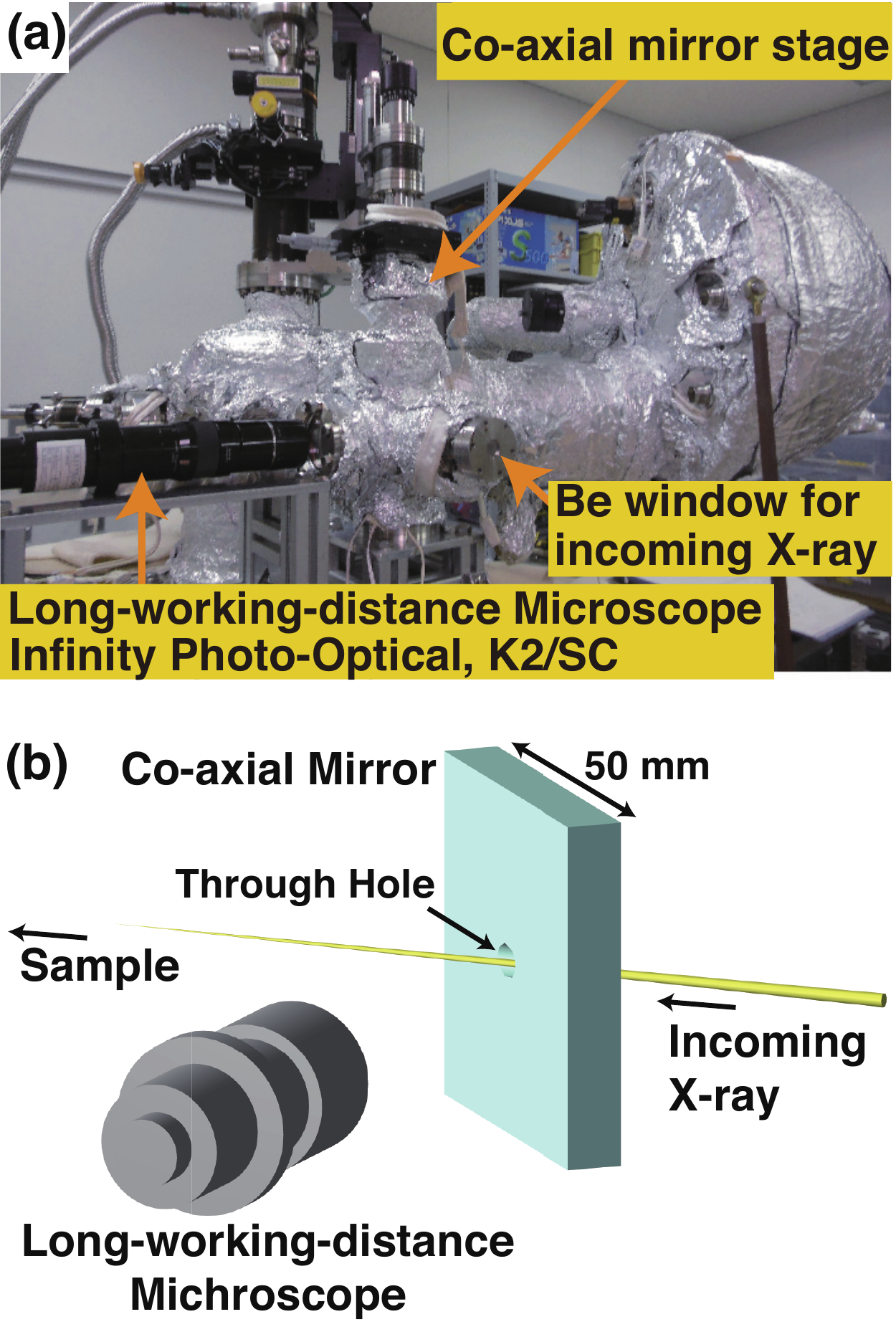}
\caption 
{(color online) Photograph of the co-axial sample monitoring system combined the long working-distance microscope with co-axial mirror in (a), and its schematic view in (b). The top view of the optical geometry is shown in Fig.~\ref{fig1}.}
\label{fig5}
\end{center}
\end{figure}

%------------\section{Fig6_K2_IMAGE}----------------
\begin{figure}
\begin{center}
\includegraphics[width=8cm,clip]{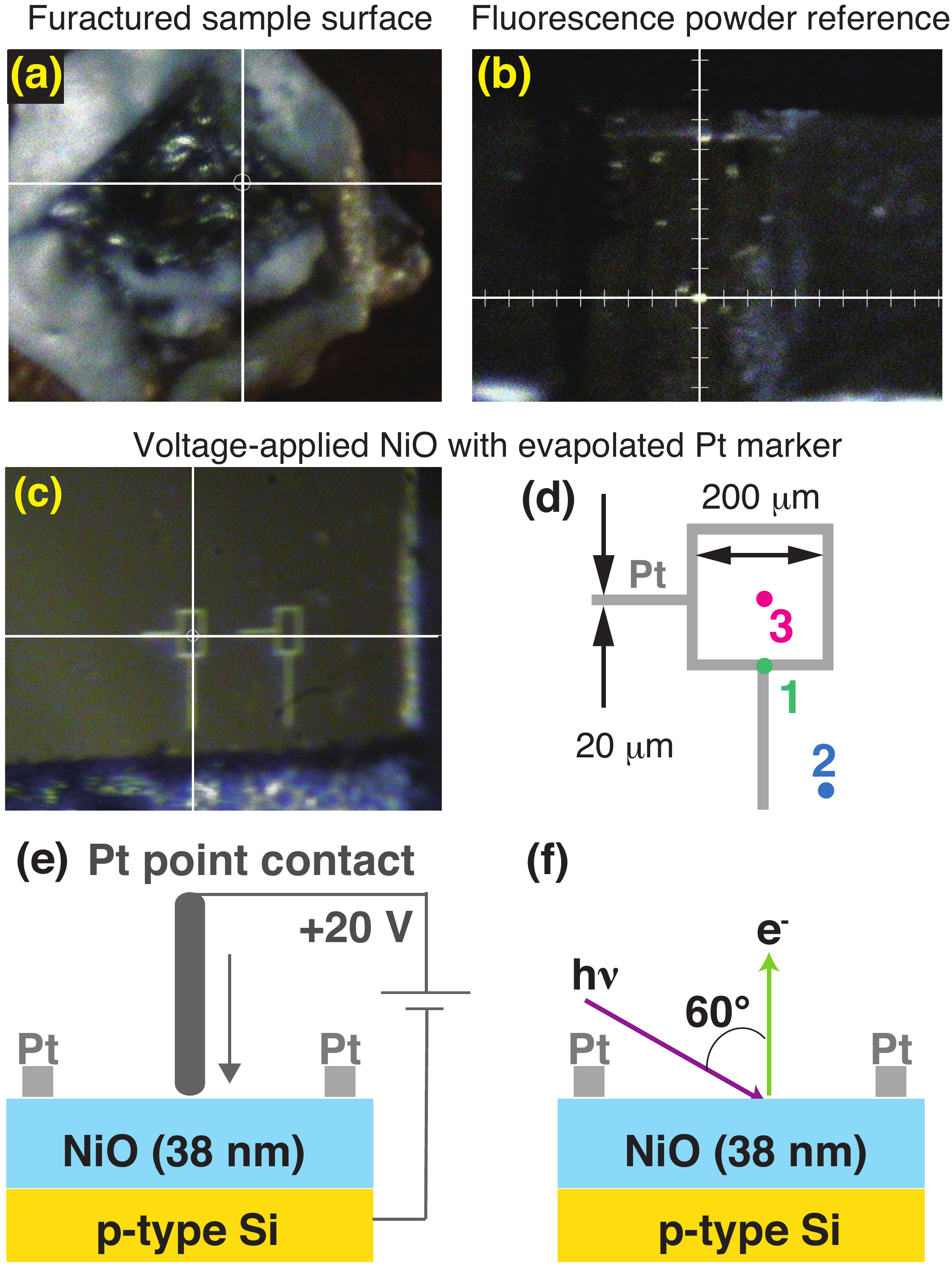}
\caption 
{(color online) Co-axial microscope image for the fractured surface of YbB$_{12}$ (a), that for the fluorescence-powder reference on the gold reference on the manipulator (b), and that for the evaporated Pt marker on NiO film (c). The X-ray beam spot is located at the intersection of the cross lines. (d) The schematic picture of the square- and line-shaped Pt marker to define specific regions on the NiO. The positions measured by HAXPES are also indicated on the Pt marker (1), as-grown NiO film (2), and the voltage-applied region on NiO (3). (e) Schematic view of the setup for applying the voltage on the NiO film by Pt point contact outside the photoemission chamber, and the experimental geometry of the HAXPES measurement in (f).}
\label{fig6}
\end{center}
\end{figure}

%------------\section{Fig7_POL.VB HAXPES}----------------
\begin{figure}
\begin{center}
\includegraphics[width=8cm,clip]{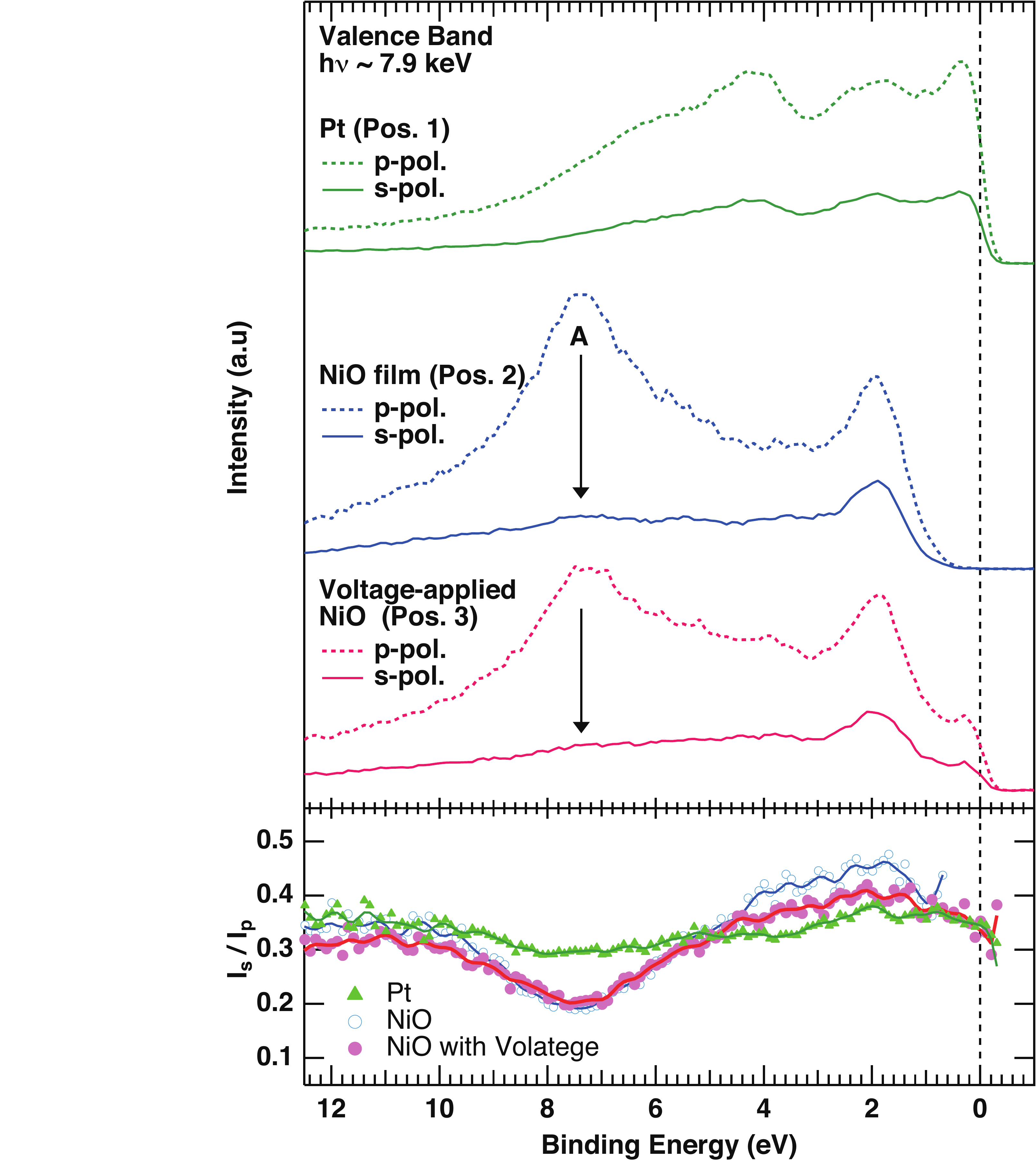}
\caption 
{(color online) Valence-band photoemission spectra of Pt measured at position 1 (pos.1) in Fig~\ref{fig6}(d), those of the as-grown NiO substrate (pos. 2) and those of the voltage-applied NiO (pos. 3) (bottom) recorded by the $p$- and $s$-polarized photons. $I_{s}$/$I_{p}$ spectra for all three positions are plotted in the lower pannel. Note that the data points ranging from 0.4 eV to Fermi level in $I_{s}$/$I_{p}$ spectrum for as-grown NiO are removed for eliminating the divergent behavior due to the gap opening.}
\label{fig7}
\end{center}
\end{figure}

%------------\section{Fig8_SUBTRACTION}----------------
\begin{figure}
\begin{center}
\includegraphics[width=8cm,clip]{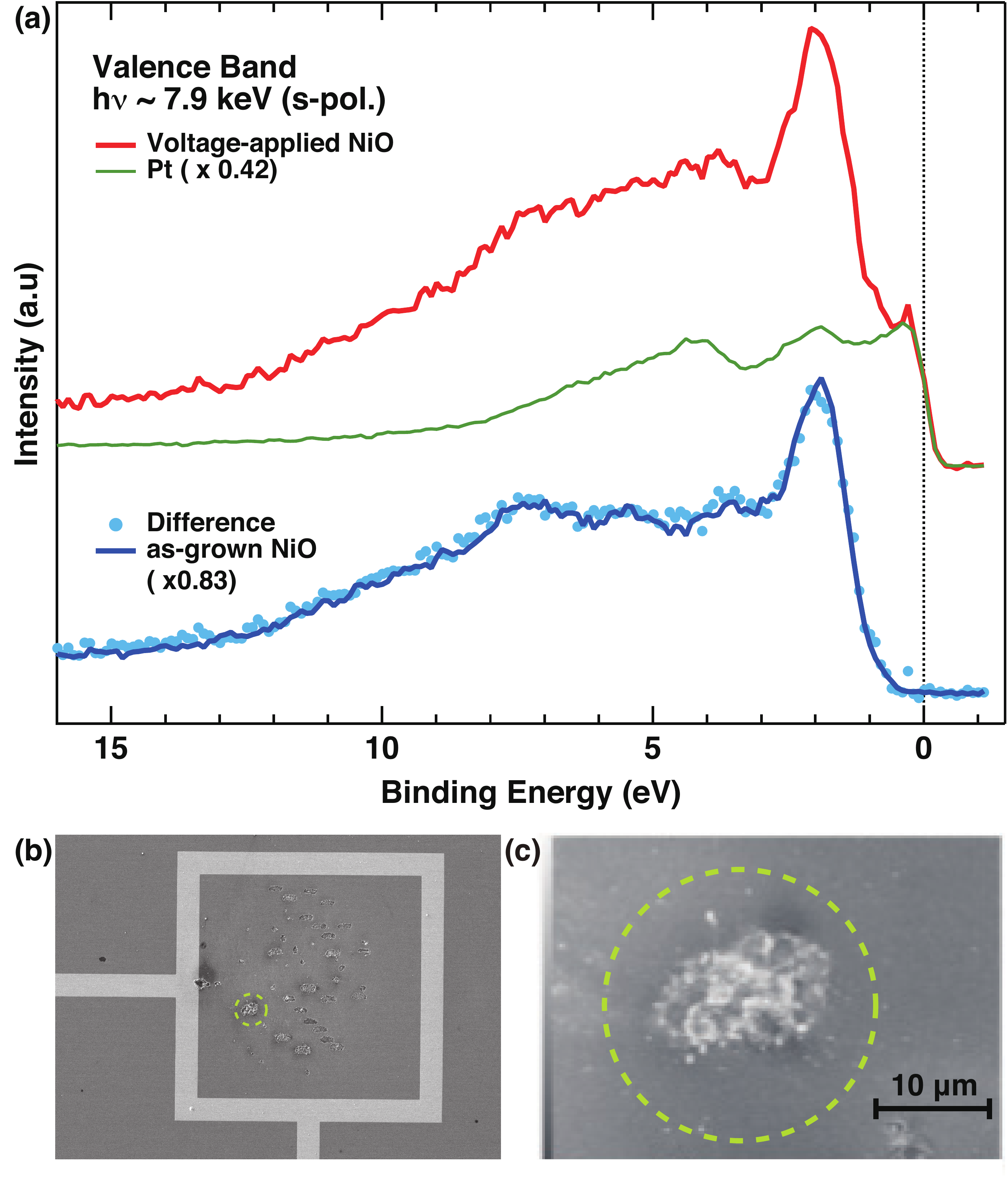}
\caption 
{(color online) (a)Valence-band photoemission spectrum with $s$-polarization geometry for the voltage-applied NiO and normalized spectrum of Pt (top). These difference is shown in bottom with the spectrum of the as-grown NiO film. (b) SEM image for voltage-applied NiO. The Dashes circle indicates one of the spots touched by Pt point contact with focused SEM image in (c).}
\label{fig8}
\end{center}
\end{figure}

The two-axis rotation is very powerful for optimizing the detection direction of the photoelectrons, but it sometimes gives a difficulty for normalizing the spectra recorded at the different angles. This might be simply due to the surface roughness; the area of the excitation X-ray on the rough surface is often not exactly the same when the sample surface is not flat, yielding the variation of the photoemission intensity. In the case of the very weak dichroic signal as shown in Fig.~\ref{fig4}, it is crucial to record the photoemission signal from the same region on the sample surface for comparison of the spectra before and after any rotations. To minimize this problem, we have also installed a co-axial sample monitoring system (Fig.~\ref{fig5}) combining the long-working-distance optical microscope (Infinity photo-optical, K2/SC) with the aluminum mirror (50 mm$^2$ square) having a 5 mm-across through-hole for the X-ray beam as shown in Fig.~\ref{fig5}(b).

As shown in Fig.~\ref{fig6}(a), the sample position is selected by the micro-positioning technique with the similar monitoring systems~\cite{Muro2009,Muro2011,Fujiwara2015}. The intersection between the incoming X-ray beam and the photoelectron analyzer axis is beforehand marked onto the microscope monitor using a fluorescent substrate, which is positioned to maximize the counts of photoelectrons detected by the photoelectron analyzer. To adjust a certain region on the sample surface to this intersection, we first set the target region to the mark on the microscope monitor, and scanned the sample position along the beam axis while keeping the target region on the mark until detecting the maximum photoelectron-counts to avoid the ambiguity of the focal depth of the microscope. Note that we have a gold reference with a small amount of phosphor powders on the aluminum finger as shown in Fig.~\ref{fig3}(b) to remark the beam spot anytime as Fig.~\ref{fig6}(b). There is no serious problem caused by the phosphor powders on the gold surface for optimizing the count rate owing to the high photoelectron kinetic energy and long probing depth at HAXPES. This simple device is helpful if one changes the magnification by changing the lens system of the microscope. 

By utilizing this positioning system, we have studied the electronic structure of NiO, which is a benchmark of the strongly correlated insulator~\cite{Sawatzky1984,Zaanen1985}, for the application as the non-volatile resistive random access memories~\cite{Gibbons1964,Kim2006,Calka2011,Jeong2012,Horiba2013}. We applied the voltage on NiO film by the Pt point contact for switching the resistivity at 50 places in the region of 200 $\mu$m $\times$ 200 $\mu$m, indicated by the square-shaped Pt marker evaporated on NiO as shown in Fig.~\ref{fig6}(c), and (d). We set the metallic states on NiO by sweeping the voltage from 0 V to +20 V between the Pt point contact touching on the NiO film and the p-type Si substrate as Fig.~\ref{fig6}(e), then release the voltage to 0 V while keeping the metallic conductivity due to the non-volatility. For studying the electronic structures of this sample by HAXPES in a different chamber by using synchrotron radiation, we have chosen the three measurement positions: Pt marker (position 1), as-grown NiO film (position 2), and the voltage-applied NiO (position 3) indicated in Fig.~\ref{fig6}(d). The closed-cycle He refrigerator was switched off for minimizing the vibration of the manipulator within $\sim$50 $\mu$m. Thus the measured temperature was set to 300 K and the overall energy resolution was set to 400 meV with the normal emission geometry as shown in Fig.~\ref{fig6}(f).

  Figure~\ref{fig7} shows the polarization-dependent valence-band HAXPES spectra recorded at these three positions. The valence-band spectra recorded on the Pt marker, namely at the position 1, show the clear Fermi cut-off, and the spectral shape is not very sensitive to the polarization~\cite{Yamasaki2014}. Meanwhile, the spectra for as-grown NiO at the position 2 outside the Pt marker clearly shows the gap opening near the Fermi level. Here clear polarization dependence of the line shape is observed. Namely, the Ni $4s$ contribution is much enhanced with the $p$-polarized geometry, especially at the position of the peak A around 7.5 eV~\cite{Sekiyama2010,Nakatsu2011,Weinen2015}. This is due to the asymmetry factors of the photoionization cross sections~\cite{Trzhaskovskaya2001,Trzhaskovskaya2002,Trzhaskovskaya2006}. We note that the spectral line shape of the valence-band spectra of the as-grown NiO is essentially the same as reported by Weinen \textit{et al.,}~\cite{Weinen2015}, in which the valence-band spectra of NiO was recorded in the complete $s$- and $p$-polarized geometries by changing the analyzer position. 

Compared with the as-grown NiO, the metallic Fermi cut-off is clearly observed at the position 3 in the voltage-applied NiO in Fig.~\ref{fig6}(d). To identify the origin of the metallic cut-off, we have compared the intensity ratio of the $s$($p$)-polarized spectra ($I_{s}$/$I_{p}$) for all three positions in the lower panel of Fig.~\ref{fig7}, for resolving the orbital contributions in the valence band spectra~\cite{Sekiyama2010}. $I_{s}$/$I_{p}$ for the voltage-applied NiO overlaps to the as-grown spectra in the range from 12 eV to 5 eV, suggesting the valence-band character is not heavily modified in this binding energy range after the application of the voltage. Meanwhile, $I_{s}$/$I_{p}$ for the voltage-applied NiO ranging from 5 eV to 1 eV is somewhat in-between those of the Pt and as-grown NiO, and overlaps fully with that for the Pt near the Fermi level, suggesting the contribution from Pt in this region on the voltage-applied NiO. We have further compared the spectral line shape of voltage-applied NiO to the Pt valence-band spectra in Fig.~\ref{fig8}. The normalized Pt valence-band spectrum so as to fit the slope of the metallic Fermi cut-off completely overlaps with the spectrum of the voltage applied NiO near the Fermi level. The subtracted spectrum of the normalized Pt spectrum from the raw spectrum of the voltages-applied NiO quantitatively reproduces the spectrum of the as-grown NiO. We should note that the Pt contribution is not due to the surrounded Pt because the length of the inner side of the marker is 200-$\mu$m, while the footprint of the X-ray beam on the sample surface is 25 ${\mu}$m $\times$ 50 ${\mu}$m (FWHM) for the normal emission geometry in which the sample normal is inclined 60 degrees to the X-ray beam as shown in Fig.~\ref{fig6}(f). Since the spatial distribution of the photoelectrons recorded by the photoelectron spectrometer was well focused, the signal from the surrounded Pt can be excluded from the spectral shape of the voltage applied NiO.

Indeed, the scanning tunneling microscope (SEM) image for the region of voltage-applied NiO shows many imprints formed by applying the voltage with Pt point contact as shown in Fig.~\ref{fig8}(b). Especially, the focused SEM image (Fig.~\ref{fig8}(c)) for one of the imprints indicated by the dashed circle in Fig.~\ref{fig8}(b) clearly shows the bump-like structure, where the energy dispersive X-ray image (not shown) revealed the fluorescence signals from Pt M-edge. Therefore, the origin of the metallic cut-off could be due to diffused Pt from the point contact during the application of the voltages for switching the resistivity. This gives an important feedback that HAXPES is not appropriate for the study of the Pt point contact method for studying the electronic structure of the NiO based resistive random access memory. Thus this result of the position-sensitive study for the electronic structure has successfully demonstrated the potential of the polarized HAXPES as a probe for revealing the driving mechanism of the device system based on the strongly correlated electron systems.

\section{CONCLUSION}

We have developed the two-axis manipulator for the polarized HAXPES, which can rotate 90 degree for azimuthal and polar angles. By designing the minimized volume and simple rotation mechanics, we have achieved the lowest cooling temperature of 9 K by a closed-cycle He refrigerator, which gives better chance to study the electronic structure of the ground states in various rare-earth compounds. To keep measuring the same sample position before and after the rotation, we have also developed the co-axial sample monitoring system with co-axial long working-distance optical microscope, supporting to record the reliable polarized HAXPES spectra to determine the ground state symmetry. We have also demonstrated the linear polarization-dependent valence-band HAXPES for the voltage-applied NiO to study the resistive switching phenomena, and found that the Pt could be diffused on NiO when one applied the voltages, forming the metallic states near the Fermi level. 

\section{ACKNOWLEDGMENTS}
This work was supported by a Grant-in-Aid for Scientific Research (23654121), Grants-in-Aid for Young Scientists (23684027, 23740240), and a Grant-in-Aid for Innovative Areas (20102003) from MEXT and JSPS, Japan, and by Toray Science Foundation. A part of sample preparation for NiO was supported by Nihon University College of Science and Technology Grants-in-Aid for Fundamental Science Research (2014). We are grateful to F. Iga for providing YbB$_{12}$ samples. HF thanks to K. Yano, S. Fujioka, H. Aratani, Y. Nakatani, and K. Terashima for supporting the experiment. The hard X-ray photoemission was performed at SPring-8 under the approval of JASRI (2014A1149, 2014B1305).

\end{document}